\documentclass[twocolumn,aps,prX,superscriptaddress]{revtex4}
\usepackage{graphicx}
\usepackage{amsmath,amssymb}
\usepackage{color}
\usepackage{bbold}
\usepackage{pdflscape}
\usepackage{float}
\usepackage{graphicx}
\usepackage{afterpage}
\usepackage{capt-of}
\usepackage[noxcolor]{pstricks}
\usepackage{pst-all}
\usepackage[color]{changebar}

\newcommand{\be}{\begin{equation}}
\newcommand{\ee}{\end{equation}}
\newcommand{\bea}{\begin{eqnarray}}
\newcommand{\eea}{\end{eqnarray}}
\newcommand{\ket} [1] {\vert #1 \rangle}

\def\ket#1{|\,#1\,\rangle}

\def\opone{\leavevmode\hbox{\small1\kern-3.8pt\normalsize1}}

\newcommand{\eins}{\mbox{$1 \hspace{-1.0mm}  {\bf l}$}}

\begin{document}

\title{Experimental Quantum Solution to the Dining Cryptographers Problem}


\author{Alley Hameedi}
\affiliation{Department of Physics, Stockholm University, S-10691,
Stockholm, Sweden}

\author{Breno Marques}
\affiliation{Instituto de F\'{i}sica, Universidade de S\~{a}o Paulo, P. O. Box 6 6318, 05315-970 S\~{a}o Paulo, Brazil}

\author{Sadiq Muhammad}
\affiliation{Department of Physics, Stockholm University, S-10691,
Stockholm, Sweden}

\author{Marcin Wie\'sniak}
\affiliation{Institute of Informatics, Faculty of Mathematics, Physics, and Informatics, University of
Gda\'{n}sk, PL-80-308 Gda\'{n}sk, Poland}


\begin{abstract}
Quantum resources such as superposition and entanglement have been used to provide unconditional key distribution, secret sharing and communication complexity reduction. In this letter we present a novel quantum information protocol for dining cryptographers problem and anonymous vote casting by a group of voters. We successfully demonstrate the experimental realization of the protocol using single photon transmission. Our implementation employs a flying particle scheme where a photon passes by the voters who perform a sequence of actions (unitary transformations) on the photonic state at their local stations. 
\end{abstract}

\author{Mohamed Bourennane}
\email{boure@fysik.su.se}
\affiliation{Department of Physics, Stockholm University, S-10691,
Stockholm, Sweden}
\maketitle
\section{INTRODUCTION}
Some of the most important features of quantum mechanics involve a system's ability to be in a superposition of two classically distinguishable states and the impossibility to perfectly distinguish between two non-orthogonal states. These features correspondingly lead to quantum information processing schemes and protocols, such as quantum cryptographic key distribution  \cite{BB84}, quantum Byzantine agreement  \cite{QBA} and quantum communication complexity reduction  \cite{QCCR}. Interestingly, these problems have found their solutions not only with the help of entangled states, commonly seen as a basic and main resource in quantum information processing, but also with sequential protocols where a single quantum system is transferred from one partner to the other. Each partner then performs an action that cumulates with others so that a final decision is reached.

In this report, we present and experimentally demonstrate a novel quantum solution to a problem of dining cryptographers \cite{CHUAM}. It was introduced by Chaum and refers to three cryptographers who went out together for a dinner in a high-class restaurant. In the end, they learned that the cheque was paid anonymously, either by one of them or by the agency they work for. The cryptographers respected each others right for secret generosity, but wanted to know if the dinner was covered by the agency. This problem is also known as the unanimous voting problem.

A simple, classical solution was proposed in the original paper and can be outlined as follows: The diners pairwise establish secret bits, $k_{ij}=0,1$, and each of them announces sum (modulo 2) of his secret bits if he did not pay or the negation of this sum if he did. Now calculating XOR of the announced sum reveals the outcome of the problem where a result 1 shows that one of them has paid, while 0 suggests that it was at the expense of the agency. 

However for a generalized case, this solution poses a number of problems and drawbacks. First is a collision, which occurs when two (or any even number) of partners have paid as their alterations would cancel each other and the agency incorrectly turns out to be the sponsor. The second is complexity, which is a complication for a large group of users. Finally, the last partner to reveal his outcome might forge the broadcast.

Note that without the collision issue, the problem of dining cryptographers is equivalent to the problem of unanimous veto voting where casting a veto is the admission of payment by a party. Typically, unanimous vetoing conforms to an organizational situation when a certain decision should be made by a group of people unanimously but the people want to keep their individual decisions a secret.

A classical solution is presented in Ref. \cite{ZIELINSKI}. In the protocol, the partners must agree on a Schnorr group and demonstrate their knowledge of the discrete logarithms in each run without revealing them. Using two broadcast rounds and the assumption that the broadcast channel available to each voter is reliable, the protocol preserves the sender's anonymity with acceptable security unless all the participants are compromised. However, as is the case with most classical solutions, this protocol relies on the difficulty in reversing of some calculations, which is an arbitrary assumption in the light of the growing available computational power.



In regards to the classical limitations, quantum solutions have also been proposed where the protocol security does not depend on the assumptions about
computational complexity. It was observed by Hillery, Ziman, Bu\v{z}ek, and Bielinkov\'{a} \cite{HILLERY} that correlations of multipartite, high-dimensional GHZ states can be used for the quantum solution. This was later rediscovered independently in \cite{RAHAMAN} where the protocol security relies on the GHZ paradox and the genuineness of the shared GHZ states between the involved parties. For a three party protocol, upon satisfying this condition, each party performs a local unitary operation on the shared GHZ state in the case of a payment. Randomly selected copies of the shared states are then used to distinguish between the cases when an even and odd number of payments are made. Once this is accomplished, the distinction between zero and double payment scenarios (for the even case) and a single against triple payment (for the odd case) is the final step in the realization of the protocol. This is possible as the GHZ states are highly symmetrical, e.g., under permutations of parties, so certain actions by one user can be compensated by others. The generalized protocol promises unconditional security along with the guarantee of no multiple payment scenarios.

Despite the complete security of the protocol, for $N$ voters, the protocol would require $(N+1)^N$-dimensional Hilbert space. Generating entangled states of many particles, even a simpler one as GHZ, is a non-trivial task and suffers from low generation rate and low state fidelity. Fortunately, the essential correlations of the GHZ states can be effectively simulated by phase shifts performed sequentially by all users on a single particle, followed by a measurement carried out by a receiver. This greatly reduces the complexity of the experiment, which now utilizes a single quantum system. 

\section{QUANTUM PROTOCOL}
Let us start with a simplified version of our protocol, which is sensitive to collisions. The protocol's feature is that it is based on a single qubit and is later generalized to a scheme without collisions.

The protocol is based  on the concept of mutually unbiased bases (MUBs). For a $d$ dimensional system, consider states $\{\{\ket{j,l}\}_{j}\}_{l=0}^{d-1}$, where $j$ enumerates the basis and $l$--the vector in the basis. $j=d$ shall always denote the computational basis and the number of other known bases depends on the dimension $d$. For powers of primes, sets of MUBs are informationally complete ($j=0,1,...,d$) \cite{WOOTTERS}. In particular, for prime $d\geq2$ we have
\begin{eqnarray}
\label{allmubs}
&\ket{d,l}=\ket{k}\quad{}\text(computational \ basis)\nonumber,\\
\forall_{j<d}&\ket{j,l}=\frac{1}{\sqrt{d}}\sum_{k=0}^{d-1}\omega^{kl+jk^2}\ket{k}.\nonumber\\
&\omega=e^{\frac{2\pi i}{d}}
\end{eqnarray}
From Eq. (\ref{allmubs}) we can observe that there are two important unitary transformations,
\begin{equation}
\label{vprime}
V_d=\text{Diag}(1,\omega,\omega^2,\omega^3,...)
\end{equation}
 cyclically permutes the vectors within any of the non-computational bases, while 
\begin{equation}
\label{uprime}
U_d=\text{Diag}(1,\omega,\omega^4,\omega^9,...)
\end{equation}
 cyclically changes the basis (neither of these operations affects the computational basis). Since the first operation transforms any state with $j\neq d$ to an orthogonal one, it will be used for casting a veto. $U_d$ on the other hand, transforms a state to a completely unbiased one, so it will be used by all participants of the protocol to improve the voter's privacy. 

\begin{figure}[http]
\begin{center}
\includegraphics[width= 9cm]{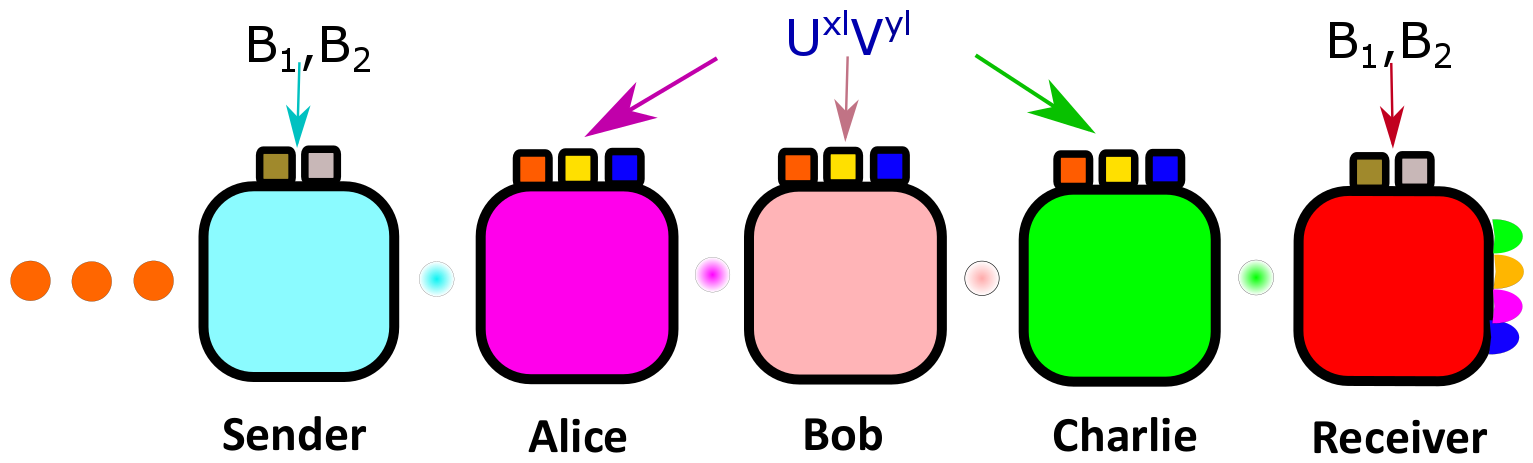}
\caption{A schematic representation of the $(N+2)$-party protocol where $N=3$. S prepares a random state, which is then propagated through $N$ voters that perform a unitary transformation ($U^{x_l}V^{y_l}$ ; $x_l,y_l\in\lbrace0,1\rbrace$). Upon arrival of a particle, the receiver performs a measurement in a randomly chosen basis.}
\label{Fig.1}
\end{center}
\end{figure}

$d=2$ is a special case, where standard formulae (\ref{allmubs}) for MUBs fail, and we have
\begin{eqnarray}
\label{dim2}
&V_2=\left(\begin{array}{cc}1&0\\0&-1\end{array}\right),\nonumber\\
&U_2=\left(\begin{array}{cc}1&0\\0&i\end{array}\right).
\end{eqnarray}
Note that $U_2^2=V_2$ so the $U_2$ not only changes the bases, but also the cycles through the states 
The basic version of the protocol for $N$ voters will involve $N+2$ parties: a sender (S), $N$ voters and a receiver (R). S has a state preparation source, R has a measurement device and the voters are provided with sets of phase shifters. This is a flying particle protocol where voters do not perform any measurement but rather execute local unitary transformations on a particle propagating among them (as shown in the Fig. $\ref{Fig.1}$ for three participants). Such schemes have found applications in experimental realizations of quantum secret sharing \cite{SECSHAR}, communication complexity reduction \cite{COMCOMP} and Byzantine agreement \cite{BYZA}.

First consider a protocol for an arbitrary number of voters $N>2$ sequentially communicating a single qubit. We begin with establishing an infrastructure to protect the voter's privacy. The voters will privately decide about the power of $U_2$ applied by them, so that an auxiliary eavesdropper will not know in which basis she should measure the state, regardless of the place of its interception. However, to make it possible for R and S to conclude the voting result, the overall transformation of this kind should be null.
This is done as follows: each voter sets his or her device to perform $U_2^{x_l}$, where $0\leq x_l\leq 3$ is a secret random integer, and $l$ labels the voter. S sends in a random state from one of the MUBs $\left\{\{\ket{0,0},\ket{0,1}\}=\frac{1}{\sqrt{2}}(\ket{0}+\ket{1}),\frac{1}{\sqrt{2}}(\ket{0}-\ket{1})\right\}$,\\$\left\{\{\ket{1,0},\ket{1,1}\}=\frac{1}{\sqrt{2}}(\ket{0}+i\ket{1}),\frac{1}{\sqrt{2}}(\ket{0}-i\ket{1})\right\}$ and R measures in one, randomly chosen, of these bases. They then discuss their choices and if they have a strong match, they accept the choice of the secret numbers, otherwise the voters choose them again. If $\sum_{l=1}^Nx_l=0\text{ mod } 4$, the overall effect of $U_2$s is the unity. On the other hand, it is unknown, to which the distributed state belongs. 

For the voting step, the sender prepares $\ket{0,0}$ or $\ket{1,0}$, chosen at random. Each voter chooses $U_2^{x_l}V_2$ if he wants to veto or $U_2^{x_l}$ otherwise. The receiver measures in one of the bases with $j=0$ or 1. If S has sent $\ket{j,0}$ and R registers $\ket{j,1}$ he can be sure that some voter has put a veto. However, registering $\ket{j,0}$ might mean any even number of vetoes, including 0. This protocol is subject to collisions and works under the assumption that at most one voter poses a veto. 

What are the ingredients of the protocol with arbitrarily many vetoes? First, there should be no collisions, so the decision should always be conclusive. The easiest way to guarantee the conclusiveness to $N$ parties is to count the votes modulo $N'>N$. The other aspect is that we may prefer not to reveal how many vetoes were posed; the only information to be known is if there were any.

In general, $N'$ can be taken as $p$, a prime larger than $N$. That way, we are certain, that there exist bases, between which the voters may switch, and within which they can change states avoiding collisions, even if all of them cast a veto. 
The voters can realize $V_{p}$ and $U_{p}^{x_l}$ as given by Eqns. (\ref{vprime}) and (\ref{uprime}). In the first step, the voters will again randomly choose their secret numbers $x_l$ from $\{0,....,p-1\}$. The sender will randomly choose state from Eqns. (\ref{allmubs}) and the receiver the basis in which he would measure. They proceed until they verify that $\sum_{l=1}^px_l=0 \text{ mod } p$. Then, in the second step, the voters together with $U_p^{x_l}$ apply an additional $V$ transformation if they want to pose a veto. S sends $\ket{0,0}$ or $\ket{1,0}$ from Eq. (\ref{allmubs}). 
If R registers the same state as sent by S, it is likely that no one vetoed.

As an example, for the case of three participants where the protocol can be run using a system with $d=4$, lets consider the two following bases (with vectors given by columns) \cite{ALLMUBSUPTO6}:
\begin{eqnarray}
\label{mubs4}
B_1&=&\frac{1}{2}\left(\begin{array}{cccc}1&1&1&1\\1&i&-1&-i\\1&-1&1&-1\\1&-i&-1&i\end{array}\right),\nonumber\\
B_2&=&\frac{1}{2}\left(\begin{array}{cccc}1&1&1&-1\\1&i&-1&i\\1&-1&1&1\\-1&i&1&i\end{array}\right).
\end{eqnarray}
$U=\text{Diag}(1,1,1,-1)$ plays a role of $U_d$ and $V=\text{Diag}(1,i,-1,-i)$ will serve to cast a veto. Note that we only use two bases, so the private numbers used by the voters to protect their privacy are simply bits. Nevertheless it does not compromise the security of the protocol, since $B_1$ and $B_2$ are complementary.

One can straight-forwardly generalize these operators for squares of primes $d=p^2$ and even other power. $V_d$ is represented by $\text{Diag}(1,\omega,\omega^2,...)$ and $U_d$ is an entangling operation-for pair of subsystem it will transform an equimodular product state into a maximally entangled one, for more constituents-to a GHZ state. for example for $d=9$, $U=\text{Diag}(1,1,1,1,\omega^3,\omega^6,1,\omega^6,\omega^3)$. Starting with a state $|0,0\rangle=\frac{1}{\sqrt{d}}(1,1,...,1)$, they lead to $d$ MUBs.

One potential loophole still refers to collisions. Since a veto is anonymous, the only reason a voter to be willing to cheat is to cancel out all possible vetoes. To do this, he would need to know how many other voters (say, $t$) would pose a veto and apply $V^{-t}$. This can be deterministically successful only if the dishonest voter knows the intentions of the honest ones. Otherwise, he can apply a random power $V$ hoping to cancel out all other vetoes. A simple way to eliminate this possibility is to impose a  hardware limitation allowing the voters to pose only a single veto.

The other issue is honesty of S and R. One option is to integrate the role of S and R with the first and the last voter, respectively, and then repeat the procedure for various orders of voters. 

Alternatively, for each run of the experiment one can allow each voter to have random and private trit $y_{l,m}=0,1,2$ ($m$ labels the run of the experiment. Upon $y_{l,m}=1$, the voters perform $U^{x_l}$ according to their chosen numbers $x_l$ in attempt to establish infrastructure. For $y_{l,m}=2$ they also cast the veto if they wish to do so, and finally for $y_{l,m}=0$ they do nothing. S sends any state involved in the protocol and R measures in the random basis. Only those runs, where the voter had all their trits equal matter. First, a list of runs with $y_{l,m}=0$. If for those runs there is a perfect correlation between states sent and received, R and S can be trusted, otherwise the protocol is aborted. Then a list of runs with all $y_{l,m}=1$ is used to verify if the infrastructure was established. If this is the case, the voters find a run with all $y_{l,m}=2$ to confirm if a veto was cast. 


\section{ROADMAP FOR THE PROTOCOL}
\subsection{R and S are trusted, three voters}
\begin{enumerate}
\item{\textit{Setup stage}: S sends a four-level quantum system prepared in a random state from sets $B_1$ and $B_2$, given in (\ref{mubs4}), where the computational basis is given by $\{|0\rangle, |1\rangle,|2\rangle, |3\rangle\}$. The voters, $A,B,C$, choose if they would apply a unitary operation $U=\text{Diag}(1,1,1,-1)$ or not. R, in turn, randomly chooses to perform a measurement in basis $B_1$ or $B_2$. Afterwards, S firstly communicates the basis that was used to prepare the state and if it does not match the basis choice of R, the round is discarded. Secondly, S communicates which state was produced in this run of the experiment. If R finds a match between the produced and the measured state, this validates the voters choice of even number of $U$. However, if there are no such correlations, the round is repeated and voters have to make their choice for $U$ again.}

\item{\textit{Voting stage}: S sends the following two ququart states
\begin{subequations}
\begin{eqnarray}
\label{quartstates}
\frac{1}{2}(|0\rangle+|1\rangle+|2\rangle+|3\rangle) 
\\
\frac{1}{2}(|0\rangle+|1\rangle+|2\rangle-|3\rangle)
\end{eqnarray}
\end{subequations}
in a random order (each state is sent exactly once). Each voter receives the particle and applies $U$ if he decided so in the previous step. If he wants to pose a veto, he also applies $V=\text{Diag}(1,i,-1,-i)\}$. R randomly chooses one of the two bases to measure for both runs.}
\item{\textit{Outcome stage}: Since S changes the initial state in each run and R measures in the same basis for both runs, they will have a match in their choices for one of the two runs. Only after they reveal their actions, it will be known that which of the two runs is relevant. In this run, if the state detected by R matches the one sent by S, it is certain that no veto was cast and the parties unanimously agree that the agency has paid. If however, R detects another state from the same basis, to what was prepared by S, it is certain that a veto was made and the state measured by R gives information about the number of vetoes.}
\end{enumerate}

\subsection{R and S are not trusted by the observers}
We shall also consider the case in which S and R are likely to conspire to produce false outcomes. For example, R can give fictional outcomes to convince everybody that a veto was cast, or S can alter the input state to ensure that a correlation is not observed. The easy solution is to allow observers to randomly perform a null transformation.
\begin{enumerate}
\item{Each voter decides if he applies $U$ and, of course, if he wants to pose a veto by applying $V$.}
\item{Each voter produces a random trit. Its values will correspond to the following actions. When 0 is drawn, they perform no action on the particle. This is to test the honesty of R and S. When 1 is drawn, they perform $U$ (if they planned so), but not $V$, which allows them to test if they have chosen the number of $U$ to be even. For 2 they perform all operations they intended, which is the voting round.}
\item{S prepares randomly one of eight states from set (\ref{mubs4}), and R measures in $B_1$ or $B_2$. Then they both share the list of state they have sent/received with the voters. Up to this stage, the choices of the observes and their random trits are kept secret.}
\item{The observers reject all the runs of the experiment, where R chose to measure in the wrong basis. This is similar to the sifting procedure in cryptography.} 
\item{Afterwards, the voters announce their random trits, but not if they applied $U$ or $V$. They reject all the runs where the trits were not equal. From the remaining rounds, the rounds with three zeros imply that the parties performed no operations on the state sent by S. If R measured the same state as well then they (S and R) are honest and dishonest otherwise. 
For the runs with three ones, if the results of S and R do not match then the voters choice of $U$, that must be even, was wrong and the protocol has to be redone. Finally, for the runs with three twos, they performed all the intended operations ($\eins$, U, V, UV).} 

\item{If R measures a given state, he is unable to infer the number of vetoes from that state without knowing the state that was prepared by S. Similarly, If S chooses randomly between $B_1$ and $B_2$ bases, this leaves no scope for the voters to cheat. Thus, the honesty of all the involved parties can be established in this way. Additionally, now considering that their honesty is proven, if the state detected by R matches the one sent by S, it can be said with certainty that no veto was cast and the parties unanimously agree that the agency has paid. If however, R detected another state from the same basis, to what was prepared by S, a veto was made and the state measured by R gives information about the number of vetoes.}`

\end{enumerate}

\section{EXPERIMENTAL REALIZATION} 

\begin{figure}[http]
\begin{center}
\includegraphics[width=9 cm]{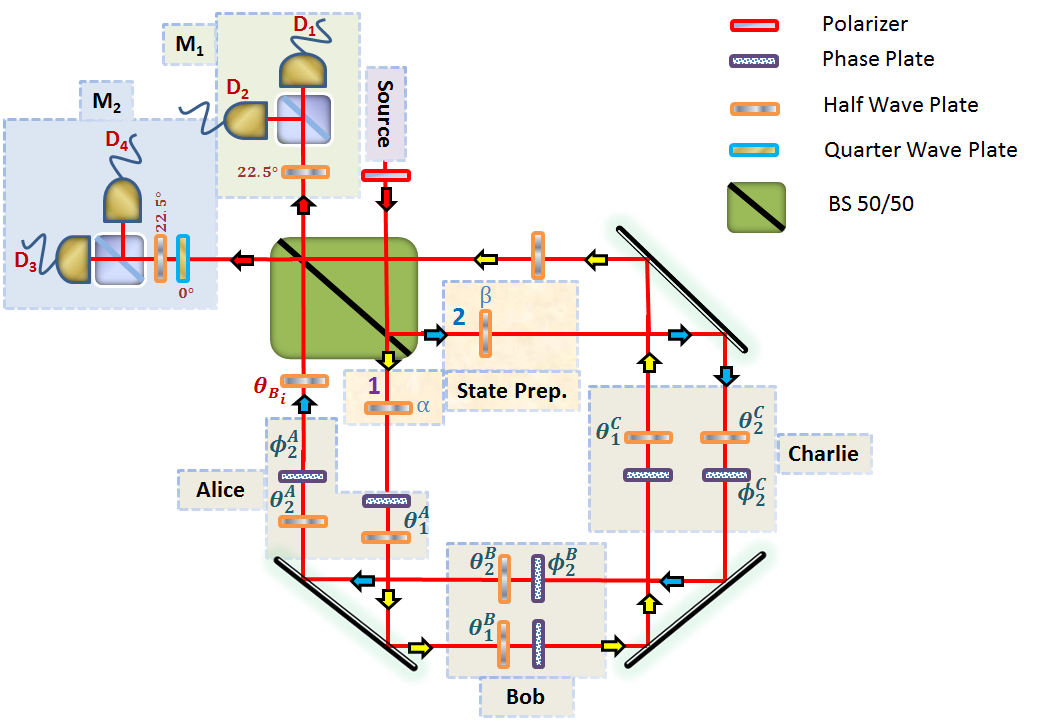}
\caption{Experimental setup. Sender S prepares states $S_{1,1},S_{1,3},S_{2,1}$ through suitable orientations of $\alpha$ and $\beta$ stationed in the two output ports (1), (2) of the BS. The three voters implement their desired transformations through suitable tilting of HWPs in path (1) and (2) and a phase shift setting PP in path (2) by mounting them on rotational stages. A HWP just before the BS in path (2) allows the receiver to choose the measurement basis. To implement the measurement, the two paths combine at the BS followed by two measurement stations $M_1$ and $M_2$.} 
\label{Fig2}
\end{center}
\end{figure}

The experimental implementation of the quantum solution for the $N+2$ party flying particle protocol is presented in Fig.~\ref{Fig2}.

Sender S prepares a four level quantum system using 2-path and 2-polarisation encoding scheme for a single photon attained through a heralded single photon source. The heralded source utilizes a spontaneous parametric down-conversion (SPDC) process where a non-linear crystal (BBO) is strongly pumped by a femtosecond pulsed laser operating at 390~nm. The SPDC process results in the creation of a pair of twin photons, commonly known as the signal and idler photons. The idler photon is then used as a trigger as its subsequent detection by a single photon detector heralds the arrival of the signal photon. These signal photons with well-defined spectral and spatial characteristics are then used as single photons for our experiment.

For a $4$-dimensional physical system, information can be encoded in the following four basis states: $|H,1\rangle$, $|V,1\rangle$, $|H,2\rangle$ and $|V,2\rangle$. ($H$) and  ($V$) are the horizontal and vertical polarizations whereas ($1$) and ($2$) are the two spatial modes of a single photon. In this way, we can write a ququart state as $a |H,1\rangle + b |V,1\rangle + c |H,2\rangle + d |V,2\rangle$.

The single photons are initially prepared in $\ket{H}$ using a polarizer oriented to horizontal polarization direction. S can prepare any state from the $B_1$ and $B_2$ bases using a 50/50 non-polarising beam splitter (BS) followed by suitably oriented half (HWP) and quarter-wave plates (QWP) alongwith a phase shift setting (PP). $\frac{1}{\sqrt{2}} \big((\cos(2\alpha) |H,1\rangle + e^{i\phi_1} \sin(2\alpha) |V,1\rangle) + e^{i\Phi}  (\cos(2\beta) |H,2\rangle + e^{i\phi_2}  \sin(2\beta) |V,2\rangle) \big).$ For the basis $B_i$, $i=\lbrace1,2\rbrace$, the states in (\ref{mubs4}) are denoted such that $S_{i,j}$ is the state defined by the column $j=\lbrace1,2,3,4\rbrace$. For a proof of principle experiment, we have chosen to prepare the following states, as an example, $S_{1,1}=(1,1,1,1)^T$ and $S_{2,1}=(1,1,1,-1)^T$ for the infrastructure establishment round and states $S_{1,1}=(1,1,1,1)^T$, $S_{1,3}=(1,-1,1,-1)^T$ for the voting round. These four states are parametrized as
\begin{equation}
\begin{split}
|\psi\rangle = \frac{1}{2} \big( & \cos(2\alpha) |H,1\rangle \pm \sin(2\alpha) |V,1\rangle \big.\\ + & \big. e^{i\Phi}  (\cos(2\beta) |H,2\rangle \pm  \sin(2\beta) |V,2\rangle) \big).
\end{split}
\label{eq6}
\end{equation}
and only a HWP ($\alpha=\beta=\pm22.5^\circ$) in each port of the BS (50:50) is sufficient for this purpose. 

The experimental setup is shown in Fig. \ref{Fig2}. For the purpose of interference stability and practicality, a Sagnac interferometer with shifted paths is implemented that allows us to manipulate each path individually.

In order to perform unitary transformations $U$, $V$ and $UV$; Alice, Bob and Charlie are each provided with two HWP oriented at $0^\circ$ in  path (1) and (2), to set the phase $\theta_1$ and $\theta_2$ between the horizontal and the vertical polarizations in each path respectively. PP in path (2) sets the required phase $\phi_2$ between the paths (1) and (2). These three plates are mounted on  rotational stages allowing the user to perform any of the desired transformation by precisely tilting them to the suitable position. The setting for $(\theta^i_1, \theta^i_2, \phi^i_2)$, where $i=(A(Alice), B(Bob), C(Charlie))$, are $(0,\pi,0)$, $(\pi/2,\pi/2,\pi)$, and $(\pi/2,-\pi/2,\pi)$ to perform $U$, $V$ and $UV$ respectively. A second PP is added in path (1) for compensation. The physical system flies through the parties before arriving at the Receiver, where it is subjected to a measurement in either of $B_1$ or $B_2$ bases. In addition, any unwanted phase arising from a component in the interferometer was carefully compensated.

The measurement part consists of a HWP in path (2) just before the BS, where the two paths recombine, followed by measurement stations $M_1$ and $M_2$ at the two output ports of the BS. The choice of the measurement basis is set by appropriately tilting the HWP to set the desired phase $\theta_{B_i}$ ($i=\lbrace1,2\rbrace$) between the polarizations in this path, similar to the voting process. For this purpose, the HWP is also mounted on a rotational stage and another HWP is used in path (1) for compensation. $M_1$ contains a HWP ($22.5^\circ$) succeeded by a polarising beam splitter (PBS). The photons arriving at each output port of the PBS are detected by multimode fiber coupled single photon detectors. $M_2$ is similar to $M_1$ but contains an additional QWP ($0^\circ$) in the beginning. This enables the photon to be projected onto any state of the chosen measurement basis represented by single photon detectors $D_i (i=1,2,3,4)$. In our scheme, the detectors $D_1$, $D_2$, $D_3$ and $D_4$ correspond to states $S_{1,1}$, $S_{1,3}$, $S_{1,4}$ and $S_{1,2}$ when $B_1$ is chosen and to states $S_{2,1}$, $S_{2,3}$ $S_{2,4}$ and $S_{2,2}$ for the basis $B_2$.

The detectors used are Silicon avalanche photodiodes (APDs) from Excelitas Technologies with an effective detection efficiency $\eta_d = 0.55$, dark count rate $R_d \simeq 400 /sec$ and a dead time of 50~ns. An FPGA based timing system is used to record the number of coincidence events between the signal and idler photons with a detection time window of $1.7$~ns. The respective probabilities are then estimated from the number of detection events at each detector. For each measurement setting, approximately $60,000$ photons were detected per second with a total measurement time of $10$~s.

For the infrastructure establishment round, S prepares randomly one state from each basis ($S_{1,1}$ and $S_{2,1}$). Alice, Bob and Charlie choose if they want to apply a $U^{x_l}$ operation or not. Bearing in mind $\sum_{l=1}^Nx_l=0 \text{ mod } 2$, the results of the corresponding measurements are presented in Table \ref{tab1}.
For the voting round, S randomly prepares one of the states ($S_{1,1}$ and $S_{1,3}$) from the basis $B_1$. If the voters choose to pose a veto, they apply $U^{x_l}V$ with $x_l\in\lbrace0,1\rbrace$. The results of the corresponding measurements are presented in Table \ref{tab2}. Table \ref{tab3} shows that when the receiver chooses the wrong measurement basis, the measurement result is inconclusive as all the detectors click with equal probabilities.

The experimental results for each measurement setting in either round are in good agreement with the expected predictions. Moreover, the interferometric visibility of our setup in all measurements is above 89$\%$. This leads to more than 93$\%$ success probability in each run of the protocol (see Table \ref{tab2}). The estimated errors include Poissonian counting statistics and systematic errors. The intrinsic imperfections of the BSs, PBSs and HWPs are the main sources of systematic errors in this experiment.

\section{CONCLUSIONS} 
The problem of dining cryptographers is a known problem in quantum communication where proposed classical solutions are not only limited but also complex in application. We have theoretically introduced and experimentally demonstrated a novel quantum solution to this problem using a one-way sequential protocol. The solution also extends to the anonymous veto problem and is simple yet highly efficient based on the distribution of the same photon among the involved parties. The obtained experimental results validate the legitimacy of the protocol while maintaining the anonymity of the involved participants. We believe that the simplicity and completeness of the protocol makes it very attractive from a quantum cryptographic perspective.

\section*{ACKNOWLEDGEMENTS}
The project was financially supported by the Knut and Alice Wallenberg foundation, Swedish research council and ERC Advanced Grant QOLAPS. B.M is supported by the grant FAPESP N$^o$ 2014/27223-2 and M.W is supported by NCN Grant no. UMO-2015/19/B/ST-2/01999.

\begin{widetext}
\newpage

\begin{table}[t]
\resizebox{\textwidth}{!}{
\begin{tabular}{|c|c|c|c|c|c|c|c|c|} \hline\cline{1-9}

$Sender$&$Alice$ & $Bob$ & $Charlie$ & $Receiver$ &   $D_1$ &  $D_2$ &  $D_3$ &  $D_4$  \\
\hline\cline{1-9}

\: $S_{1,1}$ & $\eins$ & $\eins$ & $\eins$ & $B_1$ & $0.934\pm 0.015$ & $0.006\pm0.007$  & $0.034\pm0.007$  & $0.026
\pm0.007$ \\

\: $S_{1,1}$ & $\eins$ & $\eins$ & U & $B_1$ & $0.263\pm 0.022$ & $0.232\pm0.022$  & $0.231\pm0.022$  & $0.274\pm0.022$\\

\: $S_{1,1}$ & $\eins$ & U & U & $B_1$ & $0.975\pm 0.015$ & $0.006\pm0.007$  & $0.006\pm0.007$  & $0.013\pm0.007$\\

\: $S_{1,1}$ & U & U & U & $B_1$ & $0.263\pm 0.022$ & $0.244\pm0.022$  & $0.242\pm0.022$  & $0.252\pm0.022$\\
\hline
\: $S_{1,1}$ & $\eins$ & $\eins$ & $\eins$ & $B_2$ & $0.269\pm0.022$& $0.241\pm0.022$& $0.243\pm0.022$& $ 0.247\pm0.022		$ \\
\: $S_{1,1}$ & I & U & U & $B_2$ & $0.240\pm0.022$& $0.243\pm0.022$& $0.251\pm0.022$& $0.266\pm0.022	$ \\
\hline
\: $S_{2,1}$ & $\eins$ & $\eins$ & $\eins$ & $B_1$ & $0.257\pm0.022$& $0.245\pm0.022$& $0.226\pm0.022$& $0.272\pm0.022		$\\
\: $S_{2,1}$ & I & U & U & $B_1$ & $0.251\pm0.022$& $0.243\pm0.022$& $0.242\pm0.022$& $0.265\pm0.022$ \\
\hline
\: $S_{2,1}$ & $\eins$ & $\eins$ & $\eins$ & $B_2$ & $0.952\pm0.015$ & $0.004\pm0.007$  & $0.022\pm0.007$  & $0.022
\pm0.007$ \\

\: $S_{2,1}$ & $\eins$ & $\eins$ & U & $B_2$ & $0.251\pm0.022$ & $0.235\pm0.022$  & $0.256\pm0.022$  & $0.258
\pm0.022$\\

\: $S_{2,1}$ & I & U & U & $B_2$ & $0.983\pm0.015$ & $0.004\pm0.007$  & $0.002\pm0.007$  & $0.011
\pm0.007$\\

\: $S_{2,1}$ & U & U & U & $B_2$ & $0.240\pm0.022$ & $0.261\pm0.022$  & $0.246\pm0.022$  & $0.253
\pm0.022$\\
\hline\cline{1-9}
\end{tabular}}
\caption{Infrastructure Establishment Round. The detectors $D_1$, $D_2$, $D_3$ and $D_4$ correspond to the states $S_{1,1}$, $S_{1,3}$, $S_{1,4}$ and $S_{1,2}$ when $B_1$ is chosen and to states $S_{2,1}$, $S_{2,3}$ $S_{2,4}$ and $S_{2,2}$ for the basis $B_2$. Here, the first index indicates the basis and the second indicates the corresponding state. S prepares randomly $S_{1,1}$ between $S_{2,1}$ while B measures randomly in $B1$ or $B_2$. The estimated success probabilities from detection events in each detector are also shown.  
}
\label{tab1}
\end{table}

\begin{table}
            \footnotesize
\resizebox{\textwidth}{!}{            
\begin{tabular}{|c|c|c|c|c|c|c|c|} \hline\cline{1-8}

$Sender$&$Alice$ & $Bob$ & $Charlie$ & $D_1$ &  $D_2$ &  $D_3$ &  $D_4$ \\
\hline\cline{1-8}

\: $S_{1,1}$ & $\eins$ & $\eins$ & $\eins$ &  $0.966\pm0.015$ & $0.006\pm0.007$  & $0.010\pm0.007$  & $0.018
\pm0.007$ \\\hline
\: $S_{1,1}$ & $\eins$ & $\eins$ & V & $0.016\pm0.007$  & $0.015\pm0.007$  & $0.008
\pm0.007$ & $0.962\pm0.015$ \\\hline
\: $S_{1,1}$ & $\eins$ & V & V &  $0.011\pm0.007$  & $0.961\pm0.015$ & $0.011\pm0.007$  & $0.017\pm0.007$   \\\hline
\: $S_{1,1}$ & V & V & V &  $0.011\pm0.007$  & $0.012\pm0.007$& $0.974\pm0.015$ & $0.001\pm0.007$\\\hline
\: $S_{1,1}$ & $\eins$ & U & U &  $0.975\pm0.015$ & $0.006\pm0.007$  & $0.006\pm0.007$  & $0.013\pm0.007$ \\\hline
\: $S_{1,1}$ & V & U & U &  $0.023\pm0.007$  & $0.028\pm0.007$  & $0.006\pm0.007$ & $0.943\pm0.015$ \\\hline
\: $S_{1,1}$ & $\eins$ & U & UV &  $0.032\pm0.007$  & $0.026\pm0.007$  & $0.004\pm0.007$ & $0.938\pm0.015		$ \\\hline
\: $S_{1,1}$ & V & U & UV & $0.014\pm0.007$  & $0.94\pm0.015$ &  $0.023\pm0.007$  & $0.024\pm0.007$\\\hline
\: $S_{1,1}$ & $\eins$ & UV & UV &  $0.004\pm0.007$  & $0.961\pm0.015$ &  $0.015\pm0.007$  & $0.020\pm0.007$ \\\hline
\: $S_{1,1}$ & V & UV & UV & $0.014\pm0.007$  & $0.023\pm0.007$ & $0.956\pm0.015$ & $0.007\pm0.007$\\
\hline\cline{1-8}
\end{tabular}}
\hfill

\resizebox{\textwidth}{!}{
\begin{tabular}{|c|c|c|c|c|c|c|c|} \hline\cline{1-8}

$Sender$&$Alice$ & $Bob$ & $Charlie$ & $D_1$ &  $D_2$ &  $D_3$ &  $D_4$ \\
\hline\cline{1-8}

\: $S_{1,3}$ & $\eins$ & $\eins$ & $\eins$ & $0.003\pm0.007$  & $0.966\pm0.015$ &  $0.015\pm0.007$  & $0.017
\pm0.007$\\\hline
\: $S_{1,3}$ & $\eins$ & $\eins$ & V & $0.024\pm0.007$  & $0.030\pm0.007$ & $0.941\pm0.015$ & $0.006\pm0.007$
\\\hline
\: $S_{1,3}$ & $\eins$ & V & V &  $0.955\pm0.015$ & $0.004\pm0.007$  & $0.022\pm0.007$  & $0.019\pm0.007$ \\\hline
\: $S_{1,3}$ & V & V & V &   $0.020\pm0.007$  & $0.019\pm0.007$  & $0.015
\pm0.007$ & $0.960\pm0.015$ \\\hline
\: $S_{1,3}$ & $\eins$ & U & U &  $0.002\pm0.007$  & $0.960\pm0.015$ &  $0.022\pm0.007$  & $0.016
\pm0.007$\\\hline
\: $S_{1,3}$ & V & U & U & $0.018\pm0.007$  & $0.028\pm0.007$ & $0.946\pm0.015$ & $0.008\pm0.007$\\\hline
\: $S_{1,3}$ & $\eins$ & U & UV &   $0.018\pm0.007$  & $0.024\pm0.007$ & $0.953\pm0.015$ & $0.006\pm0.007$ \\\hline
\: $S_{1,3}$ & V & U & UV &  $0.935\pm0.015$ & $0.012\pm0.007$  & $0.026\pm0.007$ & $0.028\pm0.007$\\\hline
\: $S_{1,3}$ & $\eins$ & UV & UV &  $0.954\pm0.015$ & $0.006\pm0.007$  & $0.022\pm0.007$ & $0.018\pm0.007$ \\\hline
\: $S_{1,3}$ & V & UV & UV &  $0.015\pm0.007$  & $0.018\pm0.007$ & $0.004\pm0.007$ & $0.963\pm0.015$ \\

\hline\cline{1-8}
\end{tabular}}
\caption{Voting Round. S prepares states $S_{1,1}$ or $S_{1,3}$. The voters choose to perform a veto if they decided in the last round while R measures in $B_1$ basis. Measurements with $U^{x_l}V^{y_l}$ with $x_l,y_l\in\lbrace0,1\rbrace$ are shown. The estimated success probabilities from the detection events in each detector are shown for every run.}
\label{tab2}
\end{table}

\begin{table}[t]
\centering
\resizebox{\textwidth}{!}{
\begin{tabular}{|c|c|c|c|c|c|c|c|c|} \hline\cline{1-9}

$Sender$&$Alice$ & $Bob$ & $Charlie$ & $Receiver$ &   $D_1$ &  $D_2$ &  $D_3$ &  $D_4$  \\
\hline\cline{1-9}

\: $S_{1,1}$ & $\eins$ & $\eins$ & $\eins$ & $B_2$ & $0.257\pm0.022$& $0.255\pm0.022$& $0.247\pm0.022$& $0.241\pm0.022$ \\\hline
\: $S_{1,1}$ & $\eins$ & V & V &  $B_2$ & $0.249\pm0.022$& $0.249\pm0.022$& $0.249\pm0.022$& $0.253\pm0.022$ \\\hline
\: $S_{2,1}$ & $\eins$ & $\eins$ & $\eins$ & $B_1$ & $0.252\pm0.022$& $0.239\pm0.022$& $0.272\pm0.022$& $0.237\pm0.022$\\\hline
\: $S_{2,1}$ & $\eins$ & $\eins$ & V & $B_1$ &   $0.258\pm0.022$& $0.256\pm0.022$& $0.254\pm0.022$& $0.232\pm0.022$ \\\hline
\: $S_{2,1}$ & $\eins$ & V & V & $B_1$ &    $0.256\pm0.022$& $0.262\pm0.022$& $0.241\pm0.022$& $0.242\pm0.022$ \\\hline
\: $S_{2,1}$ & V & V & V & $B_1$ &   $0.251\pm0.022$& $0.254\pm0.022$& $0.252\pm0.022$& $0.244\pm0.022$ \\

\hline\cline{1-9}
\end{tabular}}
  \caption{Voting Round. S prepares states $S_{1,1}$ or $S_{2,1}$. The voters choose to perform a veto if they decided in the last round while R measures in either of the bases $B_1$ or $B_2$. Every time R chooses the wrong measurement basis, all detectors click with equal probabilities.}
\label{tab3}
\end{table}

\end{widetext}

\end{document}